# Mechanism of ferromagnetism enhancement in a La$_{2/3}$Sr$_{1/3}$MnO$_3$ membrane released from epitaxial strain

Takahito Takeda,[1,*] Takuma Arai,[1] Kohei Yamagami,[2] Le Duc Anh,[1,3] Masaaki Tanaka,[1,3,4] Masaki Kobayashi,[1,3] and Shinobu Ohya[1,3,4,†]

[1]*Department of Electrical Engineering and Information Systems, The University of Tokyo, 7-3-1 Hongo, Bunkyo-ku, Tokyo 113-8656, Japan*
[2]*Japan Synchrotron Radiation Research Institute (JASRI), 1-1-1 Kouto, Sayo, Hyogo, 679-5198, Japan*
[3]*Center for Spintronics Research Network, The University of Tokyo, 7-3-1 Hongo, Bunkyo-ku, Tokyo 113-8656, Japan*
[4]*Institute for Nano Quantum Information Electronics (NanoQuine), The University of Tokyo, 4-6-1 Komaba, Meguro-ku, Tokyo153-8505, Japan*

Corresponding authors: [*]takeda@cryst.t.u-toky.ac.jp, [†]ohya@cryst.t.u-toky.ac.jp

**Abstract**

Recent studies have shown that the magnetic properties of the ferromagnetic perovskite oxide La$_{2/3}$Sr$_{1/3}$MnO$_3$ (LSMO) grown on an SrTiO$_3$ (STO) substrate, such as its magnetic moment and Curie temperature, can be improved by releasing the film from the substrate. However, the microscopic origin of this enhancement is not yet well understood. In this study, we use synchrotron radiation measurements to investigate the mechanism of ferromagnetism enhancement in an LSMO membrane released from an STO substrate by dissolving a water-soluble Sr$_4$Al$_2$O$_7$ buffer layer. Using resonant photoemission spectroscopy on the as-grown LSMO film and LSMO membrane, we elucidate that the strain release from the STO substrate enhances the itinerancy of the Mn-3$d$ electrons via $p$–$d$ hybridization, and this strengthens the double-exchange interaction. The reinforcement of the double-exchange interaction, in turn, improves the ferromagnetism of LSMO.

## 1. Introduction

Perovskite oxides have attracted significant attention due to their rich properties and potential device applications [1–4]. In particular, La$_{1-x}$Sr$_x$MnO$_3$ is a promising ferromagnet due to it having a Curie temperature ($T_C$) above 300 K and exhibiting half-metallicity [5,6]. The ferromagnetism of La$_{1-x}$Sr$_x$MnO$_3$ is based on the double-exchange interaction within Mn$^{3+}$-O-Mn$^{4+}$ [7]: Mn 3$d$ electrons hop to the neighbor Mn ions through the O ion between them and align the spin direction of the Mn ions. The Mn 3$d$ electrons are responsible for the transport and magnetic properties. Some experimental studies have demonstrated the excellent performance of La$_{1-x}$Sr$_x$MnO$_3$-based spintronic devices; specifically, an extremely high tunnel magnetoresistance ratio of ~1900% has been found in La$_{2/3}$Sr$_{1/3}$MnO$_3$ (LSMO)-based magnetic tunnel junctions [8,9], and a giant spin-valve ratio of 140% has been reported in La$_{1-x}$Sr$_x$MnO$_3$-based lateral spin-MOSFET devices [10]. However, these outstanding features appear only at low temperatures since the magnetization and conductivity of La$_{1-x}$Sr$_x$MnO$_3$ are significantly reduced near interfaces with other perovskite oxides [11,12]. This interface layer is called the "dead layer."

The origin of dead layers has been attributed to various interfacial effects such as strain and oxygen octahedral rotation [13–15]. Recently, epitaxial lift-off of $La_{1-x}Sr_xMnO_3$ thin films from a substrate using a water-soluble sacrificial buffer layer has attracted substantial interest as a promising approach to solving the dead-layer problem [16–22]. Some studies have reported that the magnetization and/or $T_C$ values is improved [17,19,21,22] and that the dead layer is reduced [19,22] by the epitaxial lift-off. For device applications, it is essential to understand the mechanism of the enhancement of ferromagnetism by epitaxial lift-off.

In this study, we investigate the relationship between the magnetism and electronic structures of an as-grown LSMO film and an LSMO membrane to consider the mechanism of ferromagnetism enhancement in the LSMO membrane. Resonant photoemission spectroscopy (RPES) is used to measure the occupied Mn-$3d$ partial density of states (PDOS) [23–25], and x-ray absorption spectroscopy (XAS) is used to study the unoccupied Mn-$3d$ states [26,27]. The RPES spectra of the LSMO membrane show more itinerant features of the Mn-$3d$ electrons than those of the as-grown LSMO film. Our findings suggest that strain release as a result of the epitaxial lift-off enhances the $p$–$d$ hybridization and improves the itinerancy of the Mn-$3d$ electrons, which in turn reinforces the double-exchange interaction.

## 2. Experiments

We epitaxially grew a heterostructure composed of LSMO (30 unit cells ~12 nm) / $Sr_4Al_2O_7$ (17 nm) on an $SrTiO_3$ (STO) (001) substrate by molecular beam epitaxy [Fig. 1(a)]. The detailed growth conditions were the same as those described elsewhere [22]. To transfer the LSMO layer onto a p-Si substrate [16], we attached a polydimethylsiloxane (PDMS) film to the as-grown LSMO surface and immersed the sample in deionized water for one day. After dissolving the $Sr_4Al_2O_7$ layer, we adhered the LSMO surface to the p-Si substrate by heating it for 10 min at 70 °C. By releasing the PDMS film, the LSMO membrane was transferred onto the p-Si substrate, as shown in the microscope image of Fig. 1(b). The as-grown LSMO film is fully tensile-strained, and the in-plane lattice constants are the same as those of the STO substrate; in contrast, the LSMO membrane is epitaxial-strain-free and the lattice constants are the same as those reported for an LSMO single crystal [22]. The saturation magnetization at 5 K and the $T_C$ value of the LSMO membrane (as-grown LSMO film) are 3.2 $\mu_B$/Mn (2.0 $\mu_B$/Mn) and 323 K (312 K), respectively [22].

We performed XAS and RPES experiments at beamline BL25SU of SPring-8 [28,29]. The measurements were conducted under an ultrahigh vacuum below $2.0 \times 10^{-8}$ Pa at 77 K. The angle between the hemispherical analyzer and the incident x-rays was set at 90°. For the measurements, the sample was irradiated at an angle of 7° from its surface with a circularly polarized x-ray beam with its photon energy ($hv$) varied from 630 to 680 eV. The total energy resolution, including thermal broadening, was about 100 meV. We determined the Fermi level ($E_F$) positions in the RPES spectra by measuring the Fermi-edge position of a deposited Au film in electrical contact with each of the samples. Mn $L_3$ x-ray absorption spectroscopy (XAS) spectra were obtained in the total-electron-yield mode. Note that the transfer process does not dramatically change the electronic structure near the film surface [30].

## 3. Results and discussion

We studied the Mn-3$d$ electronic structures of the as-grown LSMO film and LSMO membrane using the XAS and RPES measurements. The XAS spectra at the Mn $L_3$ absorption edge originate from electron excitation from the Mn-2$p_{3/2}$ states to the unoccupied Mn-3$d$ states. The obtained XAS spectra for the as-grown film [right panel of Fig. 2(a)] and membrane [right panel of Fig. 2(b)] were found to have similar spectral shapes to those previously reported for LSMO films [31–35]; those shapes reflect the mixed valence states between $Mn^{3+}$ and $Mn^{4+}$.

To study the occupied Mn-3$d$ electronic structures, we measured the $hv$ dependence of the RPES spectra at the Mn $L_3$ absorption edge [left panels in Figs. 2(a) and 2(b)], where $E_B$ represents binding energy. The used $hv$ values are expressed by the horizontal broken lines drawn on the XAS spectra in the right panels of Figs. 2(a) and 2(b). Resonance enhancement of the RPES signal occurs at above $hv =$ ~639 eV, while the spectra taken at $hv = 637.5$ eV for the as-grown film and 637.4 eV for the membrane are not enhanced. Here, we refer to those spectra (taken with $hv = 637.5$ and 637.4 eV, black bold curves) as the off-spectra. The RPES spectra have peaks whose positions depend on $hv$ (shown with triangles as guides to the eye) and those whose positions do not depend on $hv$ (black dashed vertical lines). The former and latter represent the Auger components and Mn-3$d$-related states, respectively. The Mn-3$d$-related components taken at $hv = 641$ eV of the membrane are different from those of the as-grown film, which are discussed later.

To emphasize the resonantly enhanced components, we plot the second-derivative mappings of the difference spectra between the off-spectra and the RPES spectra taken at various $hv$ values, as shown in Figs. 3(a) and 3(b). The mappings show several Mn-3$d$-related components (green dashed lines) and two Auger components (red dashed lines). The Auger components, especially the component represented by the red dashed bold line, are a little more distinct in the LSMO membrane than in the as-grown film. Generally, the Auger components are observed when electrons that are excited from Mn-2$p$ states to unoccupied 3$d$ states relax to different states before compensating for the Mn-2$p$ hole formed by the excitation [36–38]. Thus, the Auger components reflect the degree of the delocalization of photoexcited Mn-3$d$ electrons. Our results suggest that the Mn-3$d$ electrons in the membrane are more itinerant than those in the as-grown film; as such, the enhancement of the itinerancy is consistent with the improvement of the conductivity by the epitaxial lift-off [16].

To strengthen the evidence for this scenario, we examined the Mn-3$d$-related states of the as-grown film [Fig. 4(a)] and membrane [Fig. 4(b)] using the difference spectra between the off-spectrum and RPES spectra (at $hv = 641$ and 642.5 eV). There are five structures, labeled $A$–$E$, assigned to the Mn-3$d$ $e_g$ band, the Mn-3$d$ $t_{2g}$ band, the O-2$p$-derived mixing state composed of O 2$p$ and Mn 3$d$, the Mn-3$d$-$t_{2g}$–O-2$p$ bonding state, and the Mn-3$d$-$e_g$–O-2$p$ bonding band [23,39]. At $hv = 642.5$ eV, the difference spectrum of the membrane has a spectral shape similar to that of the as-grown film. In contrast, the difference spectra taken at $hv = 641$ eV substantially differ in shape; the membrane has larger intensities at structures $C$ and $D$ than at structures $B$ and $E$, while the as-grown film shows the opposite trend. Since structure $C$ mainly derives from the O-2$p$ states mixed with the Mn-3$d$ states [40], the observed difference indicates that the membrane has a stronger mixing of Mn-3$d$ and O-2$p$ electrons through $p$–$d$ hybridization than the as-grown film.

To analyze the behavior of the resonance enhancement, we plot the constant-initial-state (CIS) spectrum, which is a plot of the RPES intensity at constant $E_B$ as a function of $h\nu$. Figures 5(a) and 5(b) show the CIS spectra near $E_F$ ($E_B = 0.25$ eV, see the arrows in Fig. 4) and at structure $C$ ($E_B = 3.4$ eV for the as-grown film, and $E_B = 3.5$ eV for the membrane). Since the resonance enhancement is roughly proportional to the intensity of the super Coster–Kronig (SCK) decay involved in the x-ray-absorption process [41], the CIS spectra have similar spectral shapes to the XAS spectra. In Figs. 5(a) and 5(b), the peak positions of the CIS spectra (around 640.2, 641, and 642.5 eV, see the broken vertical lines) are close to the shoulder and peak positions of the Mn $L_3$ XAS spectra (black and gray curves). The CIS spectra near $E_F$ and at structure $C$ have similar shapes to the $Mn^{3+}$ and $Mn^{4+}$ $L_3$ XAS spectra, respectively [31,42]. Thus, in the Mn $L_3$ XAS spectra, the shoulder at $h\nu = \sim 640.2$ eV and peak at $h\nu = \sim 642.5$ eV mainly originate from $Mn^{3+}$, and the shoulder at $h\nu = \sim 641$ eV is induced by $Mn^{4+}$. The $Mn^{4+}$-like feature of the CIS spectrum at structure $C$ can be attributed to the stronger O-$2p$–Mn-$3d$ hybridization of $Mn^{4+}$ than that of $Mn^{3+}$ due to the negative charge-transfer energy between $Mn^{4+}$ and $O^{2-}$ ions [43–45]. The $Mn^{3+}$-like shape of the CIS spectra near $E_F$ is explained by the PDOS near $E_F$ derived from the $e_g$ electrons of Mn. The distance between the peaks of the CIS spectrum near $E_F$ is 2.4 eV, as illustrated by the red arrow in Fig. 5(b). This value is consistent with the energy difference (2–3 eV) between the $e_{g\uparrow}$ and $t_{2g\downarrow}$-derived states, where ↑ and ↓ denote the majority and minority spins, respectively [40,46,47]. Thus, the resonant enhancements at $h\nu = \sim 640.2$ eV and 642.5 eV occur in the following processes (i) and (ii), respectively:

(i)     Mn $2p^6 3d^4 (t_{2g\uparrow}^3 e_{g\uparrow}^1) + h\nu \rightarrow$ Mn $2p^6 3d^3 (t_{2g\uparrow}^3) + e^-$ (direct photoemission)
Mn $2p^6 3d^4 (t_{2g\uparrow}^3 e_{g\uparrow}^1) + h\nu \rightarrow$ Mn $2p^5 3d^5 (t_{2g\uparrow}^3 e_{g\uparrow}^2) \rightarrow$ Mn $2p^6 3d^3 (t_{2g\uparrow}^3) + e^-$ (SCK decay)

(ii)    Mn $2p^6 3d^4 (t_{2g\uparrow}^3 e_{g\uparrow}^1) + h\nu \rightarrow$ Mn $2p^6 3d^3 (t_{2g\uparrow}^3) + e^-$ (direct photoemission)
Mn $2p^6 3d^4 (t_{2g\uparrow}^3 e_{g\uparrow}^1) + h\nu \rightarrow$ Mn $2p^5 3d^5 (t_{2g\uparrow}^3 e_{g\uparrow}^1 t_{2g\downarrow}^1) \rightarrow$ Mn $2p^6 3d^3 (t_{2g\uparrow}^3) + e^-$ (SCK decay)

The influence of strain release and the associated band-structure change are apparent in the CIS spectra near $E_F$ in Figs. 5(a) and 5(b), in which it can be seen that the membrane has sharper structures than the as-grown film. The as-grown film is under tensile strain, and the membrane has the same lattice constants as single-crystal LSMO [22]. In single-crystal LSMO, the lattice is under rhombohedral distortion and has $D_{3d}$ symmetry without the Jahan–Teller (JT) effect [5,48]. Similarly, the LSMO membrane has $D_{3d}$ symmetry, splitting the $t_{2g}$ state into the $a_{1g}$ and twofold-degenerate $e_g^\pi$ states, and the $e_g$ state remains twofold degenerate [right panel of Fig. 6(a)]. Meanwhile, for the as-grown LSMO film, due to the tensile strain (and the JT effect for $Mn^{3+}$), LSMO has $D_{4h}$ symmetry, splitting the $e_g$ state into the $a_1$ ($d_{3z^2-r^2}$) and $b_1$ ($d_{x^2-y^2}$) states, and the $t_{2g}$ state to the $e$ and $b_2$ states [left panel of Fig. 6(a)]. The CIS spectra near $E_F$ reflect the unoccupied Mn-$3d$ states. Thus, for the as-grown film, the splitting of the unoccupied states makes the spectrum broader [see Figs. 5(a) and 6(a)]. The shape of the CIS spectrum at structure $C$ for the as-grown film is relatively similar to that for the membrane. In the O-$2p$–Mn-$3d$ hybridized states, the broadness of the O-$2p$ states, whose energy scale is much larger than the band modulation by strain, probably leads to the observed insensitivity to epitaxial strain.

Due to strain release, the epitaxial lift-off shifts the $E_F$ position from the $b_1$ (and $a_1$) state for the as-grown film to the $e_g$ state for the membrane [Fig. 6(a)] [27,49,50]. Thus, in the membrane, $d_{x^2-y^2}$ and $d_{3z^2-r^2}$ electrons contribute to the double-exchange interaction through *p–d* hybridization with O-2*p* electrons, more specifically between the $d_{x^2-y^2}$ and $p_x$/$p_y$ orbitals and between the $d_{3z^2-r^2}$ and $p_z$ orbitals due to the orbital symmetry. In contrast, in the as-grown film, the $d_{x^2-y^2}$ electrons of the $b_1$ state mainly cause the double-exchange interaction via *p–d* hybridization with the $p_x$/$p_y$ orbitals. The tensile strain increases the bonding length between Mn and O ions in the (001) plane and weakens the *p–d* hybridization, as shown in Fig. 6(b). Therefore, the epitaxial lift-off strengthens the *p–d* hybridization. The disappearance of the tensile strain in the membrane enhances the double-exchange interaction.

## 4. Summary

Using RPES, we have unveiled the mechanism of ferromagnetism enhancement in an LSMO membrane released from an STO substrate. Our RPES measurements on the as-grown LSMO film and LSMO membrane demonstrate that the release from the tensile strain caused by the STO substrate enhances the itineracy of the Mn-3*d* electrons via the strengthened *p–d* hybridization between the Mn-3*d* and O-2*p* electrons. The itinerant feature of the Mn-3*d* electrons enhances the ferromagnetism via the double-exchange interaction within $Mn^{3+}$-O-$Mn^{4+}$ in the LSMO membrane.

## Acknowledgments

This work was supported partly by Grants-in-Aid for Scientific Research (Nos. 23H03802, 22H04948, 22K18293, 20H05650), CREST (No. JPMJCR1777), ERATO (JPMJER2202) of the Japan Science and Technology Agency, Spintronics Research Network of Japan (Spin-RNJ). Supporting experiments at SPring-8 were approved by the Japan Synchrotron Radiation Research Institute (JASRI) Proposal Review Committee (Proposal No. 2023B1357).

## References

[1] K. Miyano, T. Tanaka, Y. Tomioka, and Y. Tokura, *Photoinduced Insulator-to-Metal Transition in a Perovskite Manganite*, Phys. Rev. Lett. **78**, 4257 (1997).

[2] A. J. Millis, *Lattice Effects in Magnetoresistive Manganese Perovskites*, Nature **392**, 147 (1998).

[3] P. Noël, F. Trier, L. M. Vicente Arche, J. Bréhin, D. C. Vaz, V. Garcia, S. Fusil, A. Barthélémy, L. Vila, M. Bibes, and J.-P. Attané, *Non-Volatile Electric Control of Spin–Charge Conversion in a $SrTiO_3$ Rashba System*, Nature **580**, 483 (2020).

[4] S. Kaneta-Takada, M. Kitamura, S. Arai, T. Arai, R. Okano, L. D. Anh, T. Endo, K. Horiba, H. Kumigashira, M. Kobayashi, M. Seki, H. Tabata, M. Tanaka, and S. Ohya, *Giant Spin-to-Charge Conversion at an All-Epitaxial Single-Crystal-Oxide Rashba Interface with a Strongly Correlated Metal Interlayer*, Nat. Commun. **13**, 5631 (2022).

[5] A. Urushibara, Y. Moritomo, T. Arima, A. Asamitsu, G. Kido, and Y. Tokura, *Insulator-Metal Transition and Giant Magnetoresistance in $La_{1-x}Sr_xMnO_3$*, Phys. Rev. B **51**, 14103 (1995).

[6] J.-H. Park, E. Vescovo, H.-J. Kim, C. Kwon, R. Ramesh, and T. Venkatesan, *Direct Evidence for a Half-Metallic Ferromagnet*, Nature **392**, 794 (1998).

[7] C. Zener, *Interaction between the D-Shells in the Transition Metals. II. Ferromagnetic Compounds of Manganese with Perovskite Structure*, Phys. Rev. **82**, 403 (1951).

[8] M. Bowen, M. Bibes, A. Barthélémy, J.-P. Contour, A. Anane, Y. Lemaître, and A. Fert, *Nearly Total Spin Polarization in $La_{2/3}Sr_{1/3}MnO_3$ from Tunneling Experiments*, Appl. Phys. Lett. **82**, 233 (2003).

[9] R. Werner, A. Y. Petrov, L. A. Miño, R. Kleiner, D. Koelle, and B. A. Davidson, *Improved Tunneling Magnetoresistance at Low Temperature in Manganite Junctions Grown by Molecular Beam Epitaxy*, Appl. Phys. Lett. **98**, 162505 (2011).

[10] T. Endo, S. Tsuruoka, Y. Tadano, S. Kaneta-Takada, Y. Seki, M. Kobayashi, L. D. Anh, M. Seki, H. Tabata, M. Tanaka, and S. Ohya, *Giant Spin-Valve Effect in Planar Spin Devices Using an Artificially Implemented Nanolength Mott-Insulator Region*, Adv. Mater. **35**, 2300110 (2023).

[11] X. Hong, A. Posadas, and C. H. Ahn, *Examining the Screening Limit of Field Effect Devices via the Metal-Insulator Transition*, Appl. Phys. Lett. **86**, 142501 (2005).

[12] Z. Liao, F. Li, P. Gao, L. Li, J. Guo, X. Pan, R. Jin, E. W. Plummer, and J. Zhang, *Origin of the Metal-Insulator Transition in Ultrathin Films of $La_{2/3}Sr_{1/3}MnO_3$*, Phys. Rev. B **92**, 125123 (2015).

[13] Y. Konishi, Z. Fang, M. Izumi, T. Manako, M. Kasai, H. Kuwahara, M. Kawasaki, K. Terakura, and Y. Tokura, *Orbital-State-Mediated Phase-Control of Manganites*, J. Phys. Soc. Japan **68**, 3790 (1999).

[14] Z. Fang, I. V Solovyev, and K. Terakura, *Phase Diagram of Tetragonal Manganites*, Phys. Rev. Lett. **84**, 3169 (2000).

[15] A. Tebano, C. Aruta, P. G. Medaglia, F. Tozzi, G. Balestrino, A. A. Sidorenko, G. Allodi, R. De Renzi, G. Ghiringhelli, C. Dallera, L. Braicovich, and N. B. Brookes, *Strain-Induced Phase Separation in $La_{0.7}Sr_{0.3}MnO_3$ Thin Films*, Phys. Rev. B **74**, 245116 (2006).

[16] D. Lu, D. J. Baek, S. S. Hong, L. F. Kourkoutis, Y. Hikita, and H. Y. Hwang, *Synthesis of Freestanding Single-Crystal Perovskite Films and Heterostructures by Etching of Sacrificial Water-Soluble Layers*, Nat. Mater. **15**, 1255 (2016).

[17] Z. Liao and J. Zhang, *Metal-to-Insulator Transition in Ultrathin Manganite Heterostructures*, Appl. Sci. **9**, (2019).

[18] Z. Lu, J. Liu, J. Feng, X. Zheng, L. Yang, C. Ge, K. Jin, Z. Wang, and R.-W. Li, *Synthesis of Single-Crystal La$_{0.67}$Sr$_{0.33}$MnO$_3$ Freestanding Films with Different Crystal-Orientation*, APL Mater. **8**, 051105 (2020).

[19] Y. Chen, X. Yuan, S. Shan, C. Zhang, R. Liu, X. Zhang, W. Zhuang, Y. Chen, Y. Xu, R. Zhang, and X. Wang, *Significant Reduction of the Dead Layers by the Strain Release in La$_{0.7}$Sr$_{0.3}$MnO$_3$ Heterostructures*, ACS Appl. Mater. Interfaces **14**, 39673 (2022).

[20] P. Salles, R. Guzman, A. Barrera, M. Ramis, J. M. Caicedo, A. Palau, W. Zhou, and M. Coll, *On the Role of the Sr$_{3-x}$Ca$_x$Al$_2$O$_6$ Sacrificial Layer Composition in Epitaxial La$_{0.7}$Sr$_{0.3}$MnO$_3$ Membranes*, Adv. Funct. Mater. **33**, 2304059 (2023).

[21] R. Atsumi, J. Shiogai, T. Yamazaki, T. Seki, K. Ueda, and J. Matsuno, *Impact of Epitaxial Strain Relaxation on Ferromagnetism in a Freestanding La$_{2/3}$Sr$_{1/3}$MnO$_3$ Membrane*, Jpn. J. Appl. Phys. **62**, 100902 (2023).

[22] T. Arai, S. Kaneta-Takada, L. D. Anh, M. Kobayashi, M. Seki, H. Tabata, M. Tanaka, and S. Ohya, *Reduced Dead Layers and Magnetic Anisotropy Change in La$_{2/3}$Sr$_{1/3}$MnO$_3$ Membranes Released from an SrTiO$_3$ Substrate*, to be published in Appl. Phys. Lett. (2024). Doi:10.1063/5.0180288.

[23] K. Horiba, A. Chikamatsu, H. Kumigashira, M. Oshima, N. Nakagawa, M. Lippmaa, K. Ono, M. Kawasaki, and H. Koinuma, *In Vacuo Photoemission Study of Atomically Controlled La$_{1-x}$Sr$_x$MnO$_3$ Thin Films: Composition Dependence of the Electronic Structure*, Phys. Rev. B **71**, 155420 (2005).

[24] H. Kumigashira, R. Hashimoto, A. Chikamatsu, M. Oshima, T. Ohnishi, M. Lippmaa, H. Wadati, A. Fujimori, K. Ono, M. Kawasaki, and H. Koinuma, *In Situ Resonant Photoemission Characterization of La$_{0.6}$Sr$_{0.4}$MnO$_3$ Layers Buried in Insulating Perovskite Oxides*, J. Appl. Phys. **99**, 08S903 (2006).

[25] M. Kobayashi, L. D. Anh, M. Suzuki, S. Kaneta-Takada, Y. Takeda, S. Fujimori, G. Shibata, A. Tanaka, M. Tanaka, S. Ohya, and A. Fujimori, *Alternation of Magnetic Anisotropy Accompanied by Metal-Insulator Transition in Strained Ultrathin Manganite Heterostructures*, Phys. Rev. Appl. **15**, 064019 (2021).

[26] R. V Chopdekar, E. Arenholz, and Y. Suzuki, *Orientation and Thickness Dependence of Magnetization at the Interfaces of Highly Spin-Polarized Manganite Thin Films*, Phys. Rev. B **79**, 104417 (2009).

[27] C. Aruta, G. Ghiringhelli, A. Tebano, N. G. Boggio, N. B. Brookes, P. G. Medaglia, and G. Balestrino, *Strain Induced X-Ray Absorption Linear Dichroism in La$_{0.7}$Sr$_{0.3}$MnO$_3$ Thin Films*, Phys. Rev. B **73**, 235121 (2006).

[28] Y. Senba, H. Ohashi, Y. Kotani, T. Nakamura, T. Muro, T. Ohkochi, N. Tsuji, H. Kishimoto, T. Miura, M. Tanaka, M. Higashiyama, S. Takahashi, Y. Ishizawa, T. Matsushita, Y. Furukawa, T. Ohata, N. Nariyama, K. Takeshita, T. Kinoshita, A. Fujiwara, M. Takata, and S. Goto, *Upgrade of Beamline BL25SU for Soft X-Ray Imaging and Spectroscopy of Solid Using*

*Nano- and Micro-Focused Beams at SPring-8*, AIP Conf. Proc. **1741**, 030044 (2016).

[29] T. Muro, Y. Senba, H. Ohashi, T. Ohkochi, T. Matsushita, T. Kinoshita, and S. Shin, *Soft X-Ray ARPES for Three-Dimensional Crystals in the Micrometre Region*, J. Synchrotron Radiat. **28**, 1631 (2021).

[30] C.-C. Chiu, Y.-W. Chang, Y.-C. Shao, Y.-C. Liu, J.-M. Lee, S.-W. Huang, W. Yang, J. Guo, F. M. F. de Groot, J.-C. Yang, and Y.-D. Chuang, *Spectroscopic Characterization of Electronic Structures of Ultra-Thin Single Crystal $La_{0.7}Sr_{0.3}MnO_3$*, Sci. Rep. **11**, 5250 (2021).

[31] G. Shibata, K. Yoshimatsu, E. Sakai, V. R. Singh, V. K. Verma, K. Ishigami, T. Harano, T. Kadono, Y. Takeda, T. Okane, Y. Saitoh, H. Yamagami, A. Sawa, H. Kumigashira, M. Oshima, T. Koide, and A. Fujimori, *Thickness-Dependent Ferromagnetic Metal to Paramagnetic Insulator Transition in $La_{0.6}Sr_{0.4}MnO_3$ Thin Films Studied by x-Ray Magnetic Circular Dichroism*, Phys. Rev. B **89**, 235123 (2014).

[32] T. Koide, H. Miyauchi, J. Okamoto, T. Shidara, T. Sekine, T. Saitoh, A. Fujimori, H. Fukutani, M. Takano, and Y. Takeda, *Close Correlation between the Magnetic Moments, Lattice Distortions, and Hybridization in $LaMnO_3$ and $La_{1-x}Sr_xMnO_{3+\delta}$: Doping-Dependent Magnetic Cir*, Phys. Rev. Lett. **87**, 246404 (2001).

[33] Z. Xu, S. Hu, R. Wu, J.-O. Wang, T. Wu, and L. Chen, *Strain-Enhanced Charge Transfer and Magnetism at a Manganite/Nickelate Interface*, ACS Appl. Mater. Interfaces **10**, 30803 (2018).

[34] Y. Kim, S. Ryu, and H. Jeen, *Strain-Effected Physical Properties of Ferromagnetic Insulating $La_{0.88}Sr_{0.12}MnO_3$ Thin Films*, RSC Adv. **9**, 2645 (2019).

[35] C. Aruta, G. Ghiringhelli, V. Bisogni, L. Braicovich, N. B. Brookes, A. Tebano, and G. Balestrino, *Orbital Occupation, Atomic Moments, and Magnetic Ordering at Interfaces of Manganite Thin Films*, Phys. Rev. B **80**, 014431 (2009).

[36] L. Sangaletti, S. Pagliara, F. Parmigiani, A. Goldoni, L. Floreano, A. Morgante, and V. Aguekian, *Giant Resonant Photoemission at the Mn 2p3d Absorption Threshold of $Cd_{1-x}Mn_xTe$*, Phys. Rev. B **67**, 233201 (2003).

[37] M. Zangrando, E. Magnano, A. Nicolaou, E. Carleschi, and F. Parmigiani, *Resonant Photoemission Spectroscopy of Thick Mn Films on Si(111) at the 2p Edge: Detection of the Two-Hole Valence-Band Satellite of Mn*, Phys. Rev. B **75**, 233402 (2007).

[38] T. Kono, M. Kakoki, T. Yoshikawa, X. Wang, K. Sumida, K. Miyamoto, T. Muro, Y. Takeda, Y. Saitoh, K. Goto, Y. Sakuraba, K. Hono, and A. Kimura, *Element-Specific Density of States of $Co_2MnGe$ Revealed by Resonant Photoelectron Spectroscopy*, Phys. Rev. B **100**, 165120 (2019).

[39] T. Hishida, K. Ohbayashi, M. Kobata, E. Ikenaga, T. Sugiyama, K. Kobayashi, M. Okawa, and T. Saitoh, *Empirical Relationship between X-Ray*

*Photoemission Spectra and Electrical Conductivity in a Colossal Magnetoresistive Manganite $La_{1-x}Sr_xMnO_3$*, J. Appl. Phys. **113**, 233702 (2013).

[40] A. Hariki, A. Yamanaka, and T. Uozumi, *Orbital- and Spin-Order Sensitive Nonlocal Screening in Mn 2p X-Ray Photoemission of $La_{1-x}Sr_xMnO_3$*, Europhys. Lett. **114**, 27003 (2016).

[41] K. C. Prince, V. R. Dhanak, P. Finetti, J. F. Walsh, R. Davis, C. A. Muryn, H. S. Dhariwal, G. Thornton, and G. van der Laan, *2p Resonant Photoemission Study of $TiO_2$S*, Phys. Rev. B **55**, 9520 (1997).

[42] J.-S. Lee, D. A. Arena, P. Yu, C. S. Nelson, R. Fan, C. J. Kinane, S. Langridge, M. D. Rossell, R. Ramesh, and C.-C. Kao, *Hidden Magnetic Configuration in Epitaxial $La_{1-x}Sr_xMnO_3$ Films*, Phys. Rev. Lett. **105**, 257204 (2010).

[43] A. E. Bocquet, T. Mizokawa, T. Saitoh, H. Namatame, and A. Fujimori, *Electronic Structure of 3d-Transition-Metal Compounds by Analysis of the 2p Core-Level Photoemission Spectra*, Phys. Rev. B **46**, 3771 (1992).

[44] D. Asakura, Y. Nanba, E. Hosono, M. Okubo, H. Niwa, H. Kiuchi, J. Miyawaki, and Y. Harada, *Mn 2p Resonant X-Ray Emission Clarifies the Redox Reaction and Charge-Transfer Effects in $LiMn_2O_4$*, Phys. Chem. Chem. Phys. **21**, 18363 (2019).

[45] T. Saitoh, A. E. Bocquet, T. Mizokawa, and A. Fujimori, *Systematic Variation of the Electronic Structure of 3d Transition-Metal Compounds*, Phys. Rev. B **52**, 7934 (1995).

[46] V. Ferrari, J. M. Pruneda, and E. Artacho, *Density Functionals and Half-Metallicity in $La_{2/3}Sr_{1/3}MnO_3$*, Phys. Status Solidi **203**, 1437 (2006).

[47] B. Zheng and N. Binggeli, *Effects of Chemical Order and Atomic Relaxation on the Electronic and Magnetic Properties of $La_{2/3}Sr_{1/3}MnO_3$*, J. Phys. Condens. Matter **21**, 115602 (2009).

[48] L. Martín-Carrón, A. de Andrés, M. J. Martínez-Lope, M. T. Casais, and J. A. Alonso, *Raman Phonons as a Probe of Disorder, Fluctuations, and Local Structure in Doped and Undoped Orthorhombic and Rhombohedral Manganites*, Phys. Rev. B **66**, 174303 (2002).

[49] G. Shibata, M. Kitamura, M. Minohara, K. Yoshimatsu, T. Kadono, K. Ishigami, T. Harano, Y. Takahashi, S. Sakamoto, Y. Nonaka, K. Ikeda, Z. Chi, M. Furuse, S. Fuchino, M. Okano, J. Fujihira, A. Uchida, K. Watanabe, H. Fujihira, S. Fujihira, A. Tanaka, H. Kumigashira, T. Koide, and A. Fujimori, *Anisotropic Spin-Density Distribution and Magnetic Anisotropy of Strained $La_{1-x}Sr_xMnO_3$ Thin Films: Angle-Dependent x-Ray Magnetic Circular Dichroism*, Npj Quantum Mater. **3**, 3 (2018).

[50] A. Galdi, C. Aruta, P. Orgiani, C. Adamo, V. Bisogni, N. B. Brookes, G. Ghiringhelli, D. G. Schlom, P. Thakur, and L. Maritato, *Electronic Band Redistribution Probed by Oxygen Absorption Spectra of $(SrMnO_3)_n(LaMnO_3)_{2n}$ Superlattices*, Phys. Rev. B **85**, 125129 (2012).

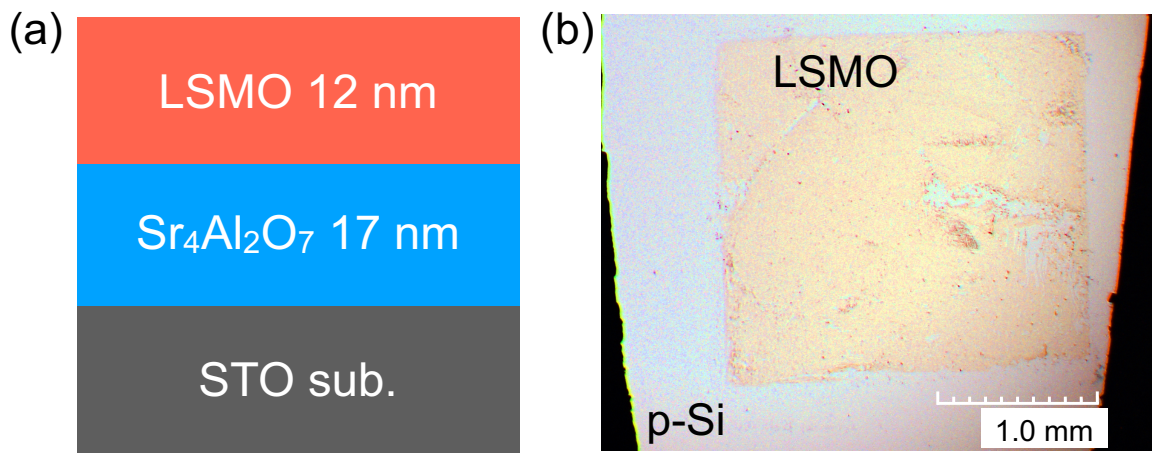

FIG. 1. Sample structures of the as-grown LSMO film and LSMO membrane. (a) Schematic of the heterostructure of the as-grown LSMO film. (b) Microscope image of the LSMO membrane transferred to a p-Si substrate.

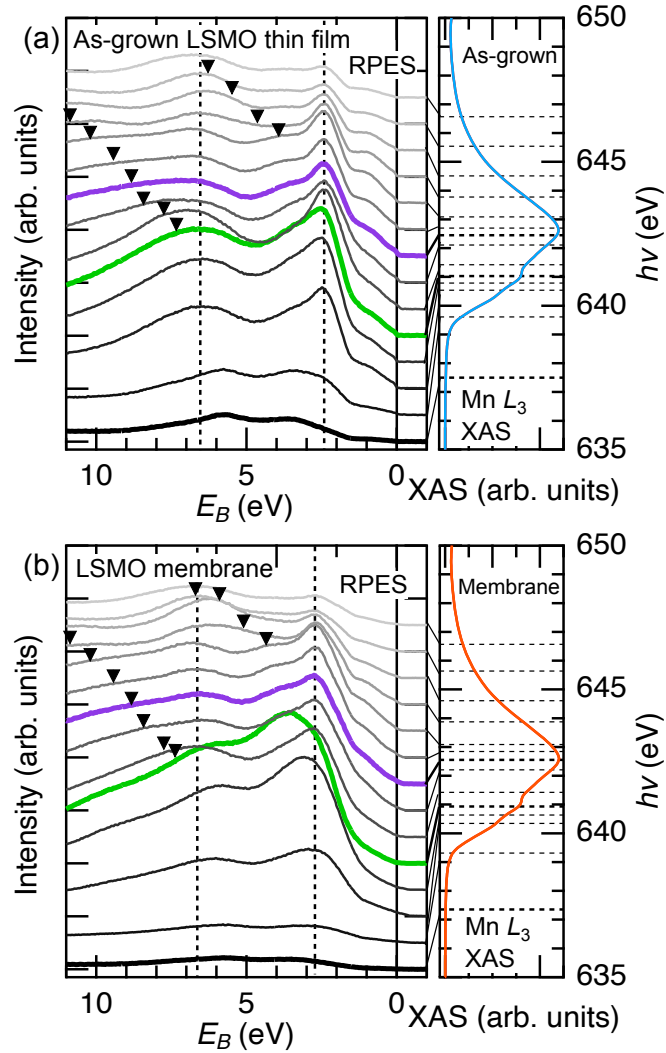

FIG. 2. RPES (left panels) and XAS (right panels) spectra at the Mn $L_3$ absorption edge of (a) the as-grown film and (b) the membrane. The vertical broken lines and triangles are guides of the Mn $3d$-related components and the two Auger components, respectively. The dotted horizontal lines on the XAS spectra of the right panels represent the $h\nu$ values that were used for measuring the RPES spectra.

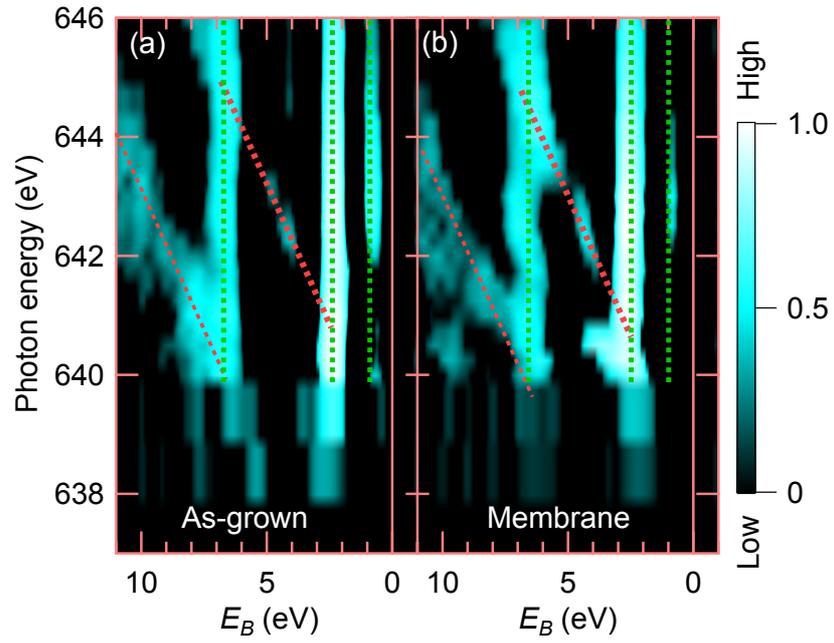

FIG. 3. Second-derivative mappings of the difference spectra between the off-spectra and RPES spectra [see Figs. 2(a) and 2(b)] for (a) the as-grown film and (b) the membrane. The dashed green and red lines represent the Mn 3*d*-related components and Auger components, respectively.

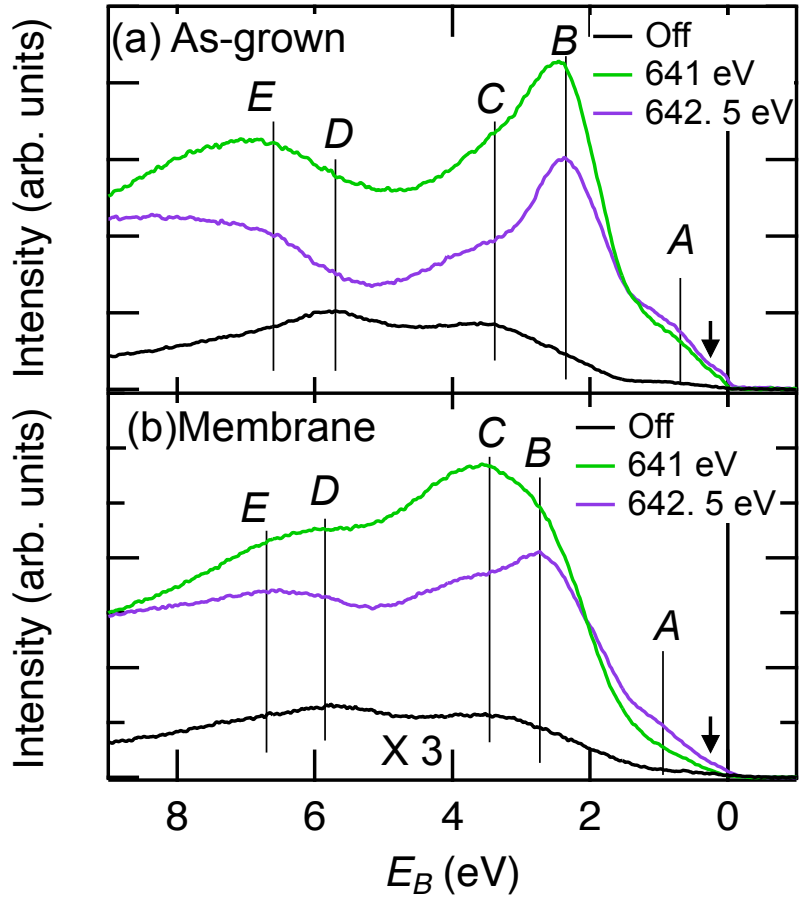

FIG. 4. Difference and off-spectra of (a) the as-grown film and (b) the membrane. The purple and green curves were obtained by subtracting the off-spectra from the purple ($hv = 642.5$ eV) and green ($hv = 641$ eV) RPES spectra in Figs. 2(a) and 2(b), respectively. The off-spectrum of the membrane is multiplied by 3.

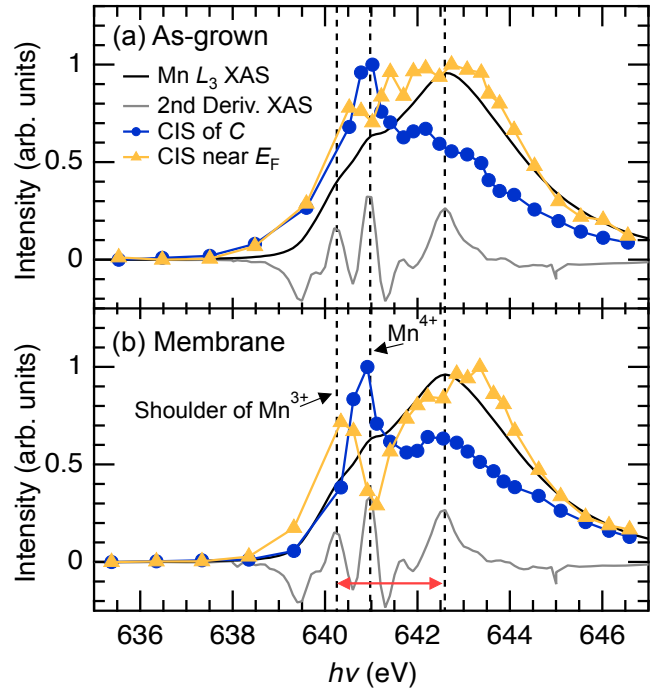

FIG. 5. Comparison between the XAS and CIS spectra at the Mn $L_3$ absorption edge of (a) the as-grown film and (b) the membrane. The blue and yellow CIS spectra were taken at structure $C$ and near $E_F$ ($E_B = 0.25$ eV), respectively. The black and gray curves represent the Mn $L_3$ XAS and the second derivative of the XAS spectra, respectively.

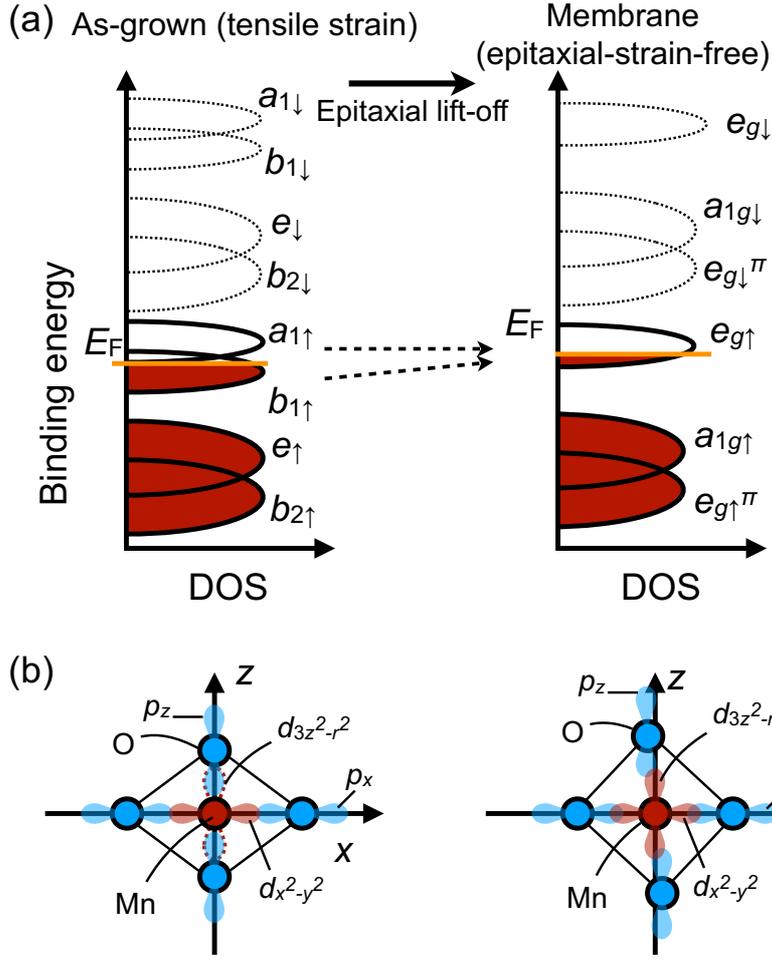

FIG. 6. Schematic illustrations of the influence of the tensile strain on (a) the Mn-3$d$ electronic structures and (b) $p$–$d$ hybridization. (a) Mn-3$d$ states of the as-grown film (tensile-strained) in the left panel and the membrane (epitaxial-strain-free) in the right panel, in which the solid and dotted black curves represent the majority- and minority-spin states, respectively. In the (epitaxial-strain-free) membrane, Mn-3$d$ states split into the $e_g$, $a_{1g}$, and $e_g^\pi$ states. In the as-grown (tensile-strained) film, the tensile strain and Jahan–Teller effect split the $e_g$ state into the $a_1$ and $b_1$ states, and the $t_{2g}$ state is similarly split into the $e$ and $b_2$ states. (b) $p$–$d$ hybridization in the $zx$ plane of tensile-strained (left panel) and epitaxial-strain-free (right panel) LSMO. The red and blue circles represent the Mn and O ions. The transparent red and blue areas represent the distribution of the Mn-3$d$ ($d_{3z^2-r^2}$, $d_{x^2-y^2}$) and O-2$p$ ($p_x$, $p_z$) orbitals, respectively. The dotted red curves represent the unoccupied $d_{3z^2-r^2}$ state. Under tensile strain, the bond lengths along the $x$ ($z$) axis are increased (decreased).

15